\documentclass[a4paper,11pt]{article}
\pdfoutput=1 

\usepackage{jheppub} 

\usepackage[T1]{fontenc} 
\usepackage[allow-number-unit-breaks, separate-uncertainty=true,range-phrase=-,range-units=single,binary-units=true]{siunitx}
\usepackage{multirow}
\usepackage[pagewise,mathlines]{lineno}

\newcommand{\nucleus}[2]{{^{#2}\text{#1}}}

\DeclareSIUnit{\wtpercent}{wt.\%}

\newcommand{\Aachen}{III. Physikalisches Institut, RWTH Aachen University, 52056 Aachen, Germany}
\newcommand{\Alabama}{Department of Physics and Astronomy, University of Alabama, Tuscaloosa, Alabama 35487, USA}
\newcommand{\Argonne}{Argonne National Laboratory, Argonne, Illinois 60439, USA}
\newcommand{\APC}{AstroParticule et Cosmologie, CNRS/IN2P3, CEA/IRFU, Observatoire de Paris, Sorbonne Paris Cit\'{e}, 75205 Paris Cedex 13, France}
\newcommand{\CBPF}{Centro Brasileiro de Pesquisas F\'{i}sicas, Rio de Janeiro, RJ, 22290-180, Brazil}
\newcommand{\CENBG}{Universit\'e de Bordeaux, CNRS/IN2P3, CENBG, F-33175 Gradignan, France}
\newcommand{\Chicago}{The Enrico Fermi Institute, The University of Chicago, Chicago, Illinois 60637, USA}
\newcommand{\CIEMAT}{Centro de Investigaciones Energ\'{e}ticas, Medioambientales y Tecnol\'{o}gicas, CIEMAT, 28040, Madrid, Spain}

\newcommand{\Drexel}{Department of Physics, Drexel University, Philadelphia, Pennsylvania 19104, USA}

\newcommand{\INR}{Institute of Nuclear Research of the Russian Academy of Sciences, Moscow 117312, Russia}
\newcommand{\CEA}{Commissariat \`{a} l'Energie Atomique et aux Energies Alternatives, Centre de Saclay, IRFU, 91191 Gif-sur-Yvette, France}

\newcommand{\Kitasato}{Department of Physics, Kitasato University, Sagamihara, 252-0373, Japan}
\newcommand{\Kobe}{Department of Physics, Kobe University, Kobe, 657-8501, Japan}
\newcommand{\Kurchatov}{NRC Kurchatov Institute, 123182 Moscow, Russia}
\newcommand{\MIT}{Massachusetts Institute of Technology, Cambridge, Massachusetts 02139, USA}
\newcommand{\MaxPlanck}{Max-Planck-Institut f\"{u}r Kernphysik, 69117 Heidelberg, Germany}
\newcommand{\NotreDame}{University of Notre Dame, Notre Dame, Indiana 46556, USA}
\newcommand{\IPHC}{IPHC, CNRS/IN2P3, Universit\'{e} de Strasbourg, 67037 Strasbourg, France}
\newcommand{\SUBATECH}{SUBATECH, CNRS/IN2P3, Universit\'{e} de Nantes, IMT-Atlantique, 44307 Nantes, France}

\newcommand{\TohokuUni}{Research Center for Neutrino Science, Tohoku University, Sendai 980-8578, Japan}

\newcommand{\TokyoInst}{Department of Physics, Tokyo Institute of Technology, Tokyo, 152-8551, Japan }
\newcommand{\TokyoMet}{Department of Physics, Tokyo Metropolitan University, Tokyo, 192-0397, Japan}
\newcommand{\TokyoSci}{Tokyo University of Science, Noda, Chiba, Japan}
\newcommand{\Muenchen}{Physik Department, Technische Universit\"{a}t M\"{u}nchen, 85748 Garching, Germany}
\newcommand{\Tubingen}{Kepler Center for Astro and Particle Physics, Universit\"{a}t T\"{u}bingen, 72076 T\"{u}bingen, Germany}

\newcommand{\UNICAMP}{Universidade Estadual de Campinas-UNICAMP, Campinas, SP, 13083-970, Brazil}
\newcommand{\vtech}{Center for Neutrino Physics, Virginia Tech, Blacksburg, Virginia 24061, USA}
\newcommand{\Chooz}{{LNCA Underground Laboratory, IN2P3/CNRS - CEA, Chooz, France}}
\newcommand{\PUC}{Pontif\'{i}cia Universidade Catolica do Rio de Janeiro, Rio de Janeiro, Brazil}
\newcommand{\Londrina}{Universidade Estadual de Londrina, 86057-970 Londrina, Brazil}
\newcommand{\Hawaii}{Physics \& Astronomy Department, University of Hawaii at Manoa, Honolulu, Hawaii, USA}
\newcommand{\IFIC}{Instituto de F\'{i}sica Corpuscular, IFIC (CSIC/UV), 46980 Paterna, Spain}
\newcommand{\KEK}{High Energy Accelerator Research Organization (KEK), Tsukuba, Ibaraki, Japan}
\newcommand{\StonyBrooks}{State University of New York at Stony Brook, Stony Brook, NY, 11755, USA}
\newcommand{\Mainz}{{Institut f\"{u}r Physik and Excellence Cluster PRISMA, Johannes Gutenberg-Universit\"{a}t Mainz, 55128 Mainz, Germany}}
\newcommand{\SK}{Kamioka Observatory, Institute for Cosmic Ray Research, University of Tokyo, Kamioka, Japan}
\newcommand{\GS}{INFN Laboratori Nazionali del Gran Sasso, 67100 Assergi (AQ), Italy}
\newcommand{\SD}{South Dakota School of Mines \& Technology, Rapid City, SD 57701}
\newcommand{\Arcadia}{Physics Department, Arcadia University, Glenside, PA 19038}

\title{\boldmath Yields and production rates of cosmogenic \texorpdfstring{$\nucleus{Li}{9}$}{9Li} and \texorpdfstring{$\nucleus{He}{8}$}{8He} measured with the Double Chooz near and far detectors }

\collaborationImg{\includegraphics[width=25mm]{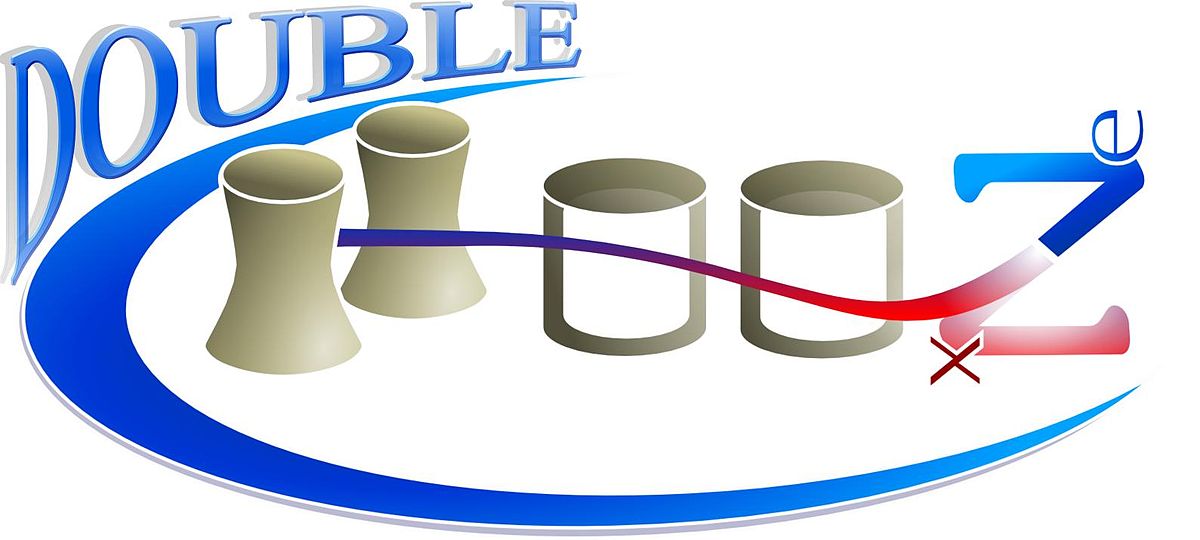}}

\author{The Double Chooz Collaboration\\}
\author[d]{H.~de~Kerret,} 
\author[e]{T.~Abrah\~{a}o,}
\author[o]{H.~Almazan,} 
\author[e]{J.C.~dos Anjos,} 
\author[v]{S.~Appel,} 
\author[k]{{J.C.~Barriere},} 
\author[a]{I.~Bekman,} 
\author[r]{T.J.C.~Bezerra,} 
\author[j]{L.~Bezrukov,} 
% ...
\emailAdd{lee.stokes@lngs.infn.it}
\emailAdd{vsibille@mit.edu}
\emailAdd{anatael@in2p3.fr}
\affiliation[a]{\Aachen} 
% ...
\affiliation[d]{\APC} 
\affiliation[e]{\CBPF} 
% ...
\affiliation[j]{\INR} 
\affiliation[k]{\CEA} 
% ...
\affiliation[o]{\MaxPlanck} 
% ...
\affiliation[r]{\SUBATECH} 
% ...
\affiliation[v]{\Muenchen} 
% ...

\keywords{neutrino detectors, neutrino experiments}

\abstract{The yields and production rates of the radioisotopes $\nucleus{Li}{9}$ and $\nucleus{He}{8}$ created by cosmic muon spallation on $\nucleus{C}{12}$, have been measured by the two detectors of the Double Chooz experiment.
The identical detectors are located at separate sites and depths, which means they are subject to different muon spectra.
The near (far) detector has an overburden of $\sim$120 m.w.e. ($\sim$300 m.w.e.) corresponding to a mean muon energy of $32.1\pm\SI{2.0}{\giga\electronvolt}$ ($63.7\pm\SI{5.5}{\giga\electronvolt}$).
Comparing the data to a detailed simulation of the $\nucleus{Li}{9}$ and $\nucleus{He}{8}$ decays, the contribution of the $\nucleus{He}{8}$ radioisotope at both detectors is found to be compatible with zero.
The observed $\nucleus{Li}{9}$ yields in the near and far detectors are $5.51\pm0.51$ and $7.90\pm0.51$, respectively, in units of $\SI{e-8}{\mu ^{-1}.g^{-1}.cm^{2} }$.
The shallow overburdens of the near and far detectors give a unique insight when combined with measurements by KamLAND and Borexino to give the first multi--experiment, data driven relationship between the $\nucleus{Li}{9}$ yield and the mean muon energy according to the power law $Y = Y_0(\left<E_\mu\right>/\SI{1}{GeV})^{\overline{\alpha}}$, giving $\overline{\alpha}=0.72\pm0.06$ and $Y_0=(0.43\pm0.11)\times \SI{e-8}{\mu ^{-1}.g^{-1}.cm^{2}}$. 
This relationship gives future liquid scintillator based experiments the ability to predict their cosmogenic $\nucleus{Li}{9}$ background rates. 
}

\begin{document}
\maketitle
\flushbottom

\section{Introduction}
\label{section:Introduction}

Cosmic muons interact with $\nucleus{C}{12}$ present in liquid scintillators creating radioisotopes through muon spallation processes, whose decays are a source of background for both neutrino and anti-neutrino experiments.
Their formation by muon capture has already been measured by Double Chooz (DC) \cite{ref:DC-mucap}.
The decays of $\nucleus{Li}{9}$ and $\nucleus{He}{8}$ are most relevant for anti-neutrinos as they can mimic the signal through their double coincidence consisting initially of the $\beta$ electron kinetic energy along with contributions from secondary particles produced through the decay and then followed by a neutron $n$--capture.
This decay signal, consisting of a prompt and then delayed event gives them the nomenclature of $\beta\text{-}n$ emitters.

Their relatively long lifetimes (\SI{257}{\milli\second} for $\nucleus{Li}{9}$ and \SI{172}{\milli\second} for $\nucleus{He}{8}$) and high muon rates at shallow detector sites means that rejecting them through a total veto applied after each muon would shrink the live time to zero.
On the other hand, the difficulty of determining their contribution to the signal makes this cosmogenic background one of the most relevant for the measurement of the neutrino mixing angle $\theta_{13}$ and any other liquid scintillator based experiments searching for anti-neutrinos, e.g the upcoming JUNO experiment \cite{ref:JUNO}.

The Near (ND) and Far (FD) Detectors are protected by distinct overburdens of $\sim 120\,\text{m.w.e}$ (metre water equivalent) and $\sim 300\,\text{m.w.e}$, respectively.
The profile and depth of the overburdens mean that each detector is subject to different muon fluxes and therefore cosmogenic background rates.
The goal of this paper is the measurement of the $\nucleus{Li}{9}$ and $\nucleus{He}{8}$ production yields at the two detector sites (i.e. at different overburdens and therefore different mean muon energies).
This begins with the generation of the expected $\nucleus{Li}{9}$ and $\nucleus{He}{8}$ spectra, described in section \ref{section:Prediction}. 
Data selection, which ultimately contains a mixture of $\nucleus{Li}{9}$ and $\nucleus{He}{8}$ events is explained in section \ref{subsec:sel_9Li_and_8He_cands} and estimation of the ND and FD $\nucleus{He}{8}$ fractions are shown in section \ref{section:Fraction}.
The $\nucleus{He}{8}$ fraction is used in section \ref{section:yields} to estimate the values or upper limits where necessary of the yields, production rates, and cross section of each cosmogenic radioisotope at the ND and FD.
As the overburdens of the DC detectors are relatively shallow, especially for the ND which started taking data after the FD, they can be combined with existing measurements by KamLAND \cite{ref:KamLANDCosmo} and Borexino \cite{ref:BorexinoCosmo} to evaluate the exponential law \cite{ref:Hagner2000} describing the production yield as a function of the mean muon energy.
This relationship will allow future liquid scintillator experiments searching for IBD signals the capability to estimate their cosmogenic background rates.

\section{The Double Chooz experiment}
\label{section:DCdet}

The Double Chooz experiment (DC) provided a measurement of the neutrino mixing angle $\theta_{13}$ by observing a deficit of anti-neutrinos created by the Chooz nuclear reactors, firstly using the far detector only~\cite{ref:DC-GdIII,ref:DC-HIII} and recently with two detectors~\cite{ref:Neutrino18} which allows most of the systematics to cancel each other out, in particular from the anti-neutrino flux.
Anti--neutrino detection is based on the Inverse Beta Decay (IBD) process ($\overline{\nu}_e + p \rightarrow e^+ + n$) in liquid scintillators, where the neutron is predominantly captured on Hydrogen (H) or Gadolinium (Gd) corresponding to released energies of \SI{2.2}{\mega\electronvolt} and $\sim$\SI{8}{\mega\electronvolt}, respectively.
This interaction is identified by a fast coincidence signal, consisting of the prompt positron signal and then the delayed $n$--capture.

The ND and FD are almost identical in construction, but situated at $\sim$\SI{400}{\metre} and $\sim$\SI{1050}{\metre} from the Chooz Nuclear Power Plant reactors, respectively.
They are made of four concentric cylindrical vessels, of which, the innermost, the Neutrino Target (NT) is filled with $\SI{10.3}{m^{3}}$ of Gd--loaded (\SI{0.1}{\wtpercent}) liquid scintillator.
The NT is surrounded by $\SI{22.5}{m^{3}}$ of Gd--free liquid scintillator called the Gamma Catcher (GC).
The NT and GC vessels are made of transparent acrylic and together they form the fiducial volume used for the detection of IBDs and the cosmogenic $\beta\text{-}n$ emitters.
Gadolinium has a larger cross section for neutron capture than H, so the majority of neutrons in the IBDs in the NT are captured on Gd, whilst in the GC there is no Gd so all the captures are on H.
With its Gd--loading, the NT was meant to act as the detection volume for anti-neutrinos, but with novel background reduction techniques \cite{ref:DC-HIII} the fiducial volume could be widened to include the GC.
The liquid scintillator in the NT is composed of ortho--phenylxylylethane (o--PXE)/n--dodecane mixed in a volume ratio of 20/80, giving a $\nucleus{C}{12}$ density of $\SI{4.31e28}{\tonne^{-1}}$.
The GC contains \SI{66}{\percent} mineral oil, \SI{30}{\percent} of n--dodecane, and \SI{4}{\percent} o--PXE, giving a $\nucleus{C}{12}$ density of $\SI{4.29e28}{\tonne^{-1}}$.

\begin{figure}[tbp]
\centering
\includegraphics[width=0.7\textwidth]{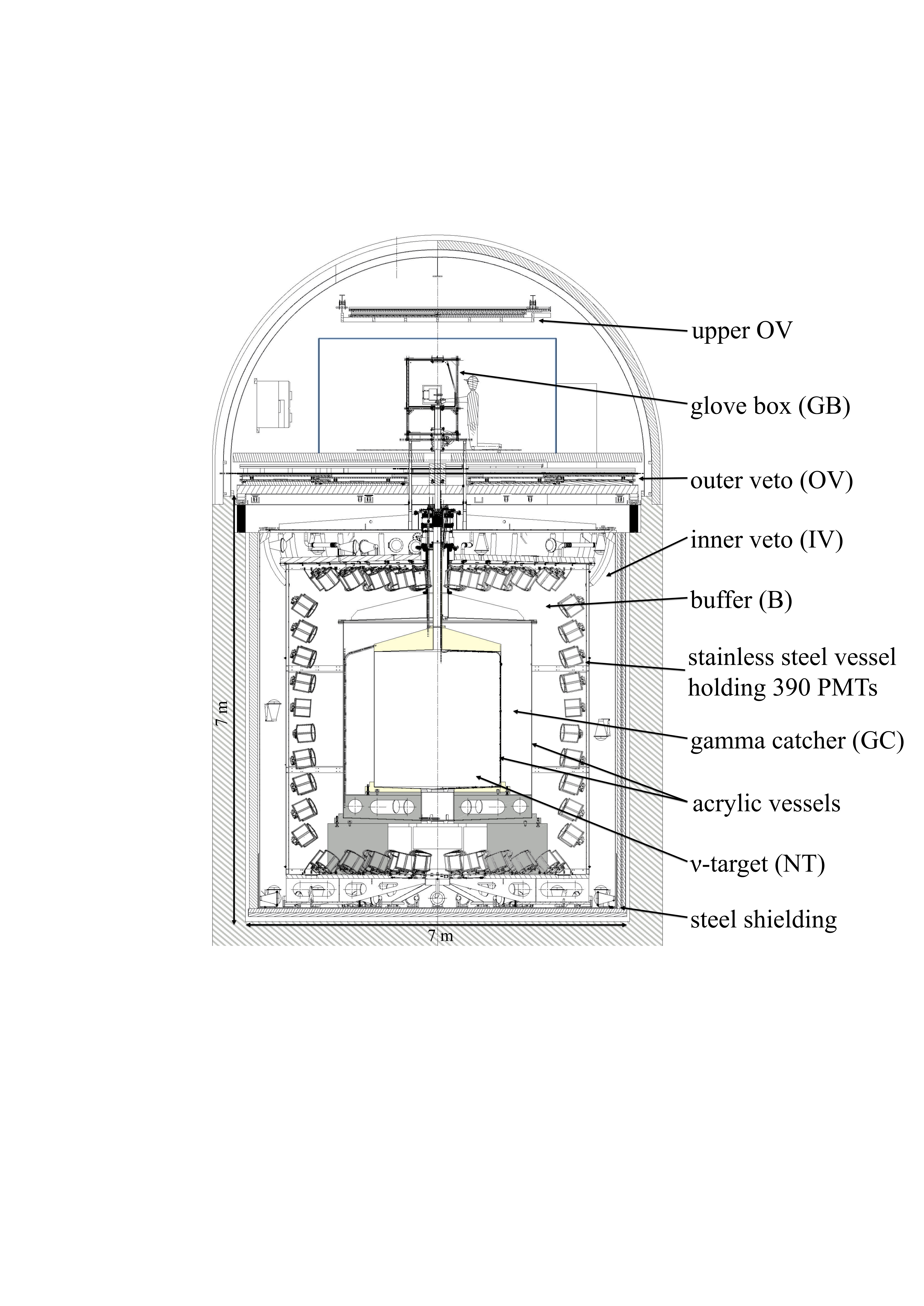}
\caption{\label{fig:detector} Schematic of the Double Chooz detectors.}
\end{figure}

The buffer encompasses the GC and is filled with non--scintillating mineral oil, acting as a shield from the surrounding radioactivity.
Detection of the scintillation light is made using 390 photomultiplier tubes (PMTs) housed within the buffer and attached to the surrounding stainless steel tank.
These three inner regions are collectively called the Inner Detector (ID), outside of which is the Inner Veto (IV), filled with liquid scintillator and 78 PMTs used to detect cosmic muons and as a shield to external radiation.
The ID and IV are then surrounded by another shield composed of \SI{1}{\metre} of water in the ND and \SI{15}{\centi\metre} of steel in the FD.
Finally, the top of the detector is covered by plastic scintillator strips called the Outer muon Veto (OV).
A schematic of the detector is shown in figure \ref{fig:detector} and more information can be found in \cite{ref:DC-GdIII,ref:DC-GdII}.

\section{Cosmogenic spectra prediction}
\label{section:Prediction}

The expected spectra of the $\nucleus{Li}{9}$ and $\nucleus{He}{8}$ decays are required to estimate their relative proportions and therefore rates produced in each detector.
As the decay schemes are intricate, the standard \textsc{Geant4} \cite{ref:geant4i,ref:geant4ii,ref:geant4iii} simulation toolkit cannot generate these spectra or provide a treatment of systematic uncertainties. Instead, each decay branch is simulated individually, the raw outgoing energies are fed into the \textsc{Geant4}--based detector simulation and dynamically combined during error estimation. In the following, a \textit{raw} spectrum is a theoretical decay spectrum of a given branch whilst a \textit{predicted} spectrum includes the detector response, analysis selection and sampling over all the decay paths.

\subsection{Raw spectra generation}
\label{subsec:raw_spectra}

The decays of the cosmogenic radioisotopes $\nucleus{Li}{9}$ and $\nucleus{He}{8}$ release various particles which all appear at the same space--time point in the liquid scintillator. Such a property stems from the large widths of all the intermediate states in the decay trees, which go hand in hand with extremely short life--times. The instant observation of all these particles by the detector defines a single \textit{prompt} event. For most decays, the energy signature of this prompt signal is dominated by the primary $\beta$--decay as contributions from heavier particles are quenched in the liquid scintillator. 

After having thermalised, the neutron makes for a \textit{delayed} event in which one or more gammas from its capture are released into the scintillator. 
The energy spectrum of the delayed events provides no handle on the $\beta\text{-}n$ emitter, so for the remainder of the paper \textit{prompt energy spectrum} is referred to as \textit{spectrum} for simplicity.

\begin{figure}[tbp]
	\centering
	\includegraphics[width=0.46\textwidth]{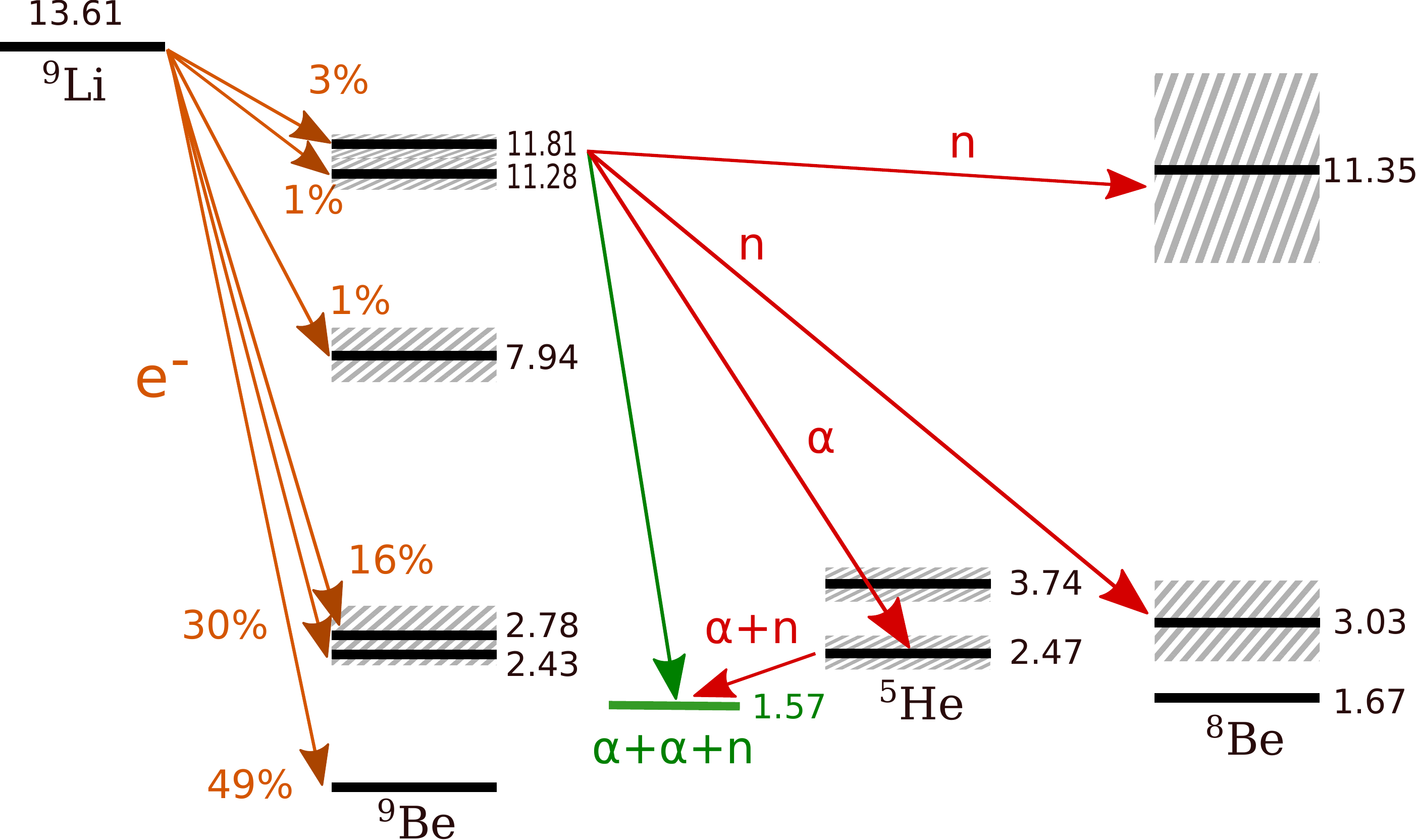}
	\hfill
	\includegraphics[width=0.52\textwidth]{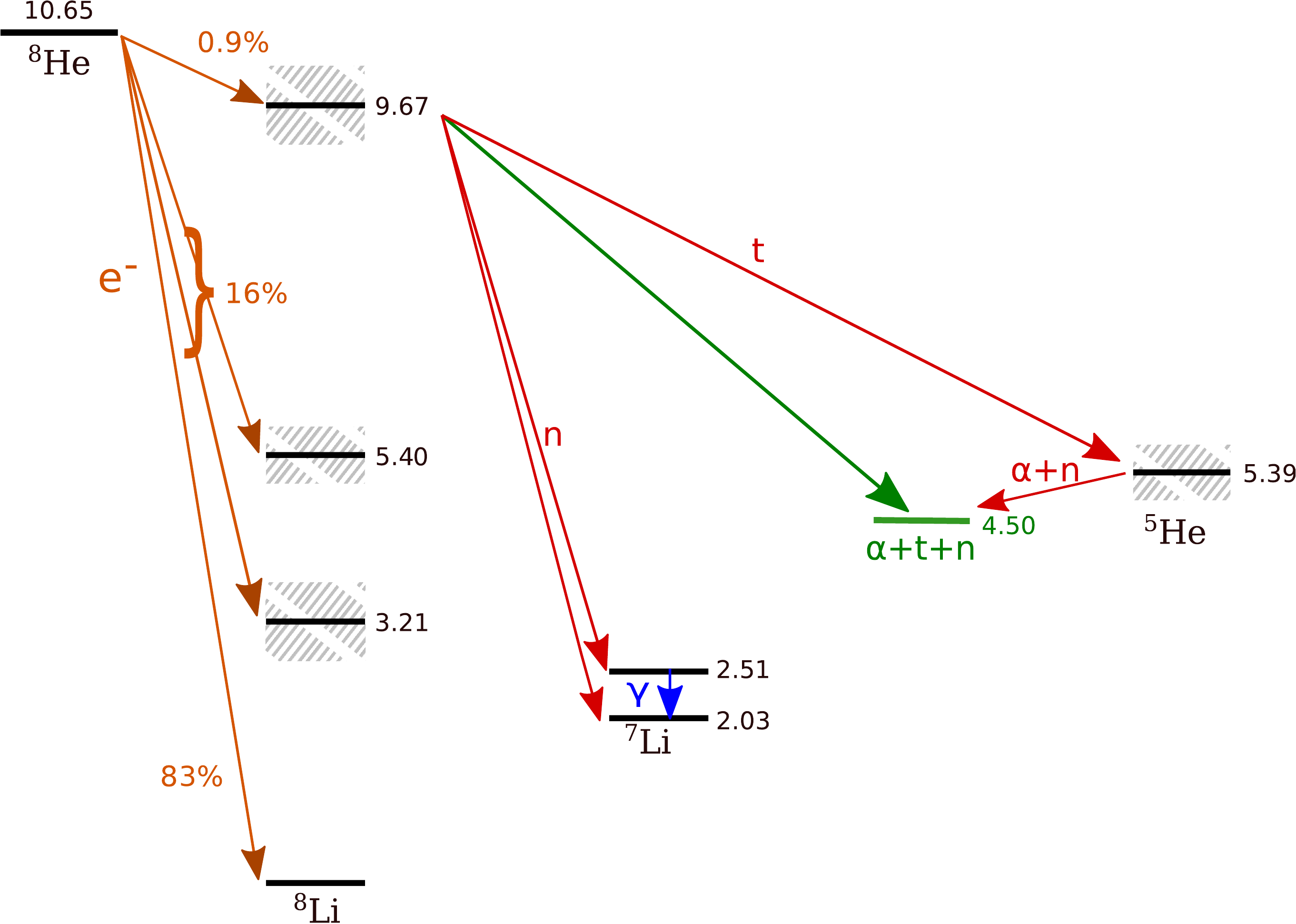}
\caption{\label{fig:lihedecays} The decay schemes for $\nucleus{Li}{9}$ (left) and $\nucleus{He}{8}$ (right), where the energy widths of the states are proportional to the hatched boxes.
The $\beta$--decay branching ratios are indicated along with some of the possible decay paths.
The particles released during a transition between levels may be found above the red arrows. The green arrows indicate a direct three-body break-up.
In both schemes, the energy levels are in $\si{\mega\electronvolt}$. 
For $\nucleus{Li}{9}$, the energy levels are quoted in relation to the $\nucleus{Be}{9}$ ground state and the final state for the $\beta\text{-}n$ decay path is always $\alpha+\alpha+n$.
For $\nucleus{He}{8}$, the energy levels are in relation to the $\nucleus{Li}{8}$ ground state. Note that the ground state of $\nucleus{Li}{7}$ is stable. }
\end{figure}

Figure \ref{fig:lihedecays} shows the $\nucleus{Li}{9}$ and $\nucleus{He}{8}$ decay schemes used in the simulation chain; the mean value of the energy levels and the $\beta ^{\text{-}}$ branching ratios are based on \cite{ref:tilley}. For clarity, only a few decay paths are shown in figure \ref{fig:lihedecays}, but all the decays which are not forbidden by kinematics have been included. A comprehensive list may be found in appendix \ref{sec:explicit_decay_chains}.
To correctly assess the energy deposited by each cosmogenic decay, predictions need to be made about the energies of all particles in the chain. 
In this work, the raw energies of the electrons, are computed using corrections to the Fermi theory for allowed $\beta$--decays. 
These corrections account for weak magnetism, radiative, and finite--size effects. 
All the other particles, namely alphas, neutrons, gammas, and tritons, are modelled using fully relativistic kinematics, correctly treating the recoil of light nuclei such as $\nucleus{He}{5}$. 
In the case of many--body break--ups, a recurrence method on the phase space --- largely inspired by \cite{ref:hagedorn} --- is implemented. 
The largeness of the state widths in the cosmogenic decay trees induces a bias on the mean energy of the states if they are generated following Lorentz distributions, as is custom with narrower widths. 
Instead, the decay chains have been modelled using Gaussian distributions.

As there is little information about the decays following the primary $\beta$--decay in either decay tree, one raw spectrum is produced per branch in the tree. 
In total, this amounts to nine spectra for $\nucleus{He}{8}$, and twenty--four for $\nucleus{Li}{9}$, all under the form of HEP \textsc{Geant4}--compatible files. 
The position of events in the ID is generated according to the $^{12}\text{C}$ density, whose value drives the cosmogenic isotope production. 

These generated, raw events are then processed through the detector simulation as described in \cite{ref:DC-GdIII}.

\subsection{Mean spectrum and covariance building}
\label{subsec:error_estimation}

The generation method presented in \ref{subsec:raw_spectra} has to be coupled with an error estimation tool to build a mean detected spectrum for each radioisotope. The branching ratios and weak magnetism corrections are subject to uncertainty, whose handling is detailed below.

The uncertainties on the branching ratios are included by varying their values between their physical bounds. When no nuclear data are available, this is achieved by uniformly selecting numbers at random so that the decay probabilities of one state add up to one. In the case of the branching ratios for the $\beta$--decays, or the strong decays of the $\SI{11.81}{\mega\electronvolt}$ level in ${\nucleus{Be}{9}}$, more constraining bounds --- retrieved from published fits to experimental data \cite{ref:tilley, ref:prezado} --- are utilised.
Every set of the branching ratios thus picked produces a possible spectrum for the considered radioisotope. 
Such a spectrum is a realisation of the multivariate random variable $\mathbf{B}$, where $\mathbf{B}$ stands for the vector of bin contents of the resulting spectrum. 
A $(n+1)$--th realisation $\mathbf{b}^{(n+1)}$ of the random variable $\mathbf{B}$, updates the estimator $\widehat{V}$ of the covariance matrix between the different bin contents as follows:
\begin{equation}
\widehat{V}_{n+1}=
\frac{1}{n}\sum_{k=1}^{n+1} 
\left(\mathbf{b}^{(k)}-\overline{\mathbf{b}}_{n+1}\right)
\left(\mathbf{b}^{(k)}-\overline{\mathbf{b}}_{n+1}\right)^{T},
\end{equation}
where $\overline{\mathbf{b}}_{n+1}$ is the sample mean after $n+1$ iterations. The size of $\widehat{V}$ is the square of the number of energy bins. 
Convergence is assumed after a few million iterations when the last fifty matrices are close to one another. 

A second matrix $W$, which represents the systematic uncertainty on the weak magnetism correction, is added to $\widehat{V}$. 
This matrix reads $W = \boldsymbol{\sigma} \boldsymbol{\sigma}^{T}$ with $\boldsymbol{\sigma}$ the vector of errors defined by $\sigma_i= C e_i B_i$, where $C$ is the slope of the error, $e_i$ the $i$--th bin centre of the histogram representing the spectrum, and $B_i$ the content of the $i$--th bin. 
The value of $C$ was set to $C= \SI{0.005}{\per\mega\electronvolt}$. This represents a conservative $100\%$ error on the value of the weak magnetism correction for allowed transitions (when compared to the linearisation of the correction found in \cite{ref:hayes}). 
It should be noted that the relative importance of this uncertainty increases with energy, as does the contribution of the electrons to the bin contents of the resulting spectrum.

Separate predicted spectra are created depending on whether the neutron is captured on Gd or H. 
This is because the energy scale is treated differently for each as they dominate different volumes of the detector, with slightly separate properties. 
As a result, the correlations between the predicted spectra had to be derived.
For each cosmogenic radioisotope, the covariance between the predicted spectra $S^{\text{Gd}}$ (obtained from an analysis of the raw spectra on Gd) and $S^{\text{H}}$ (from an analysis on H), has been computed using a variation of the aforementioned technique. 
Regardless of the capture used, the $\nucleus{Li}{9}$ and $\nucleus{He}{8}$ spectra are not correlated.

\subsection{Resulting \texorpdfstring{$\nucleus{Li}{9}$}{9Li} and \texorpdfstring{$\nucleus{He}{8}$}{8He} spectra}
\label{subsec:resulting_9Li_and_8He_spectra}

\begin{figure}[tbp]
\centering
\includegraphics[width=0.7\textwidth]{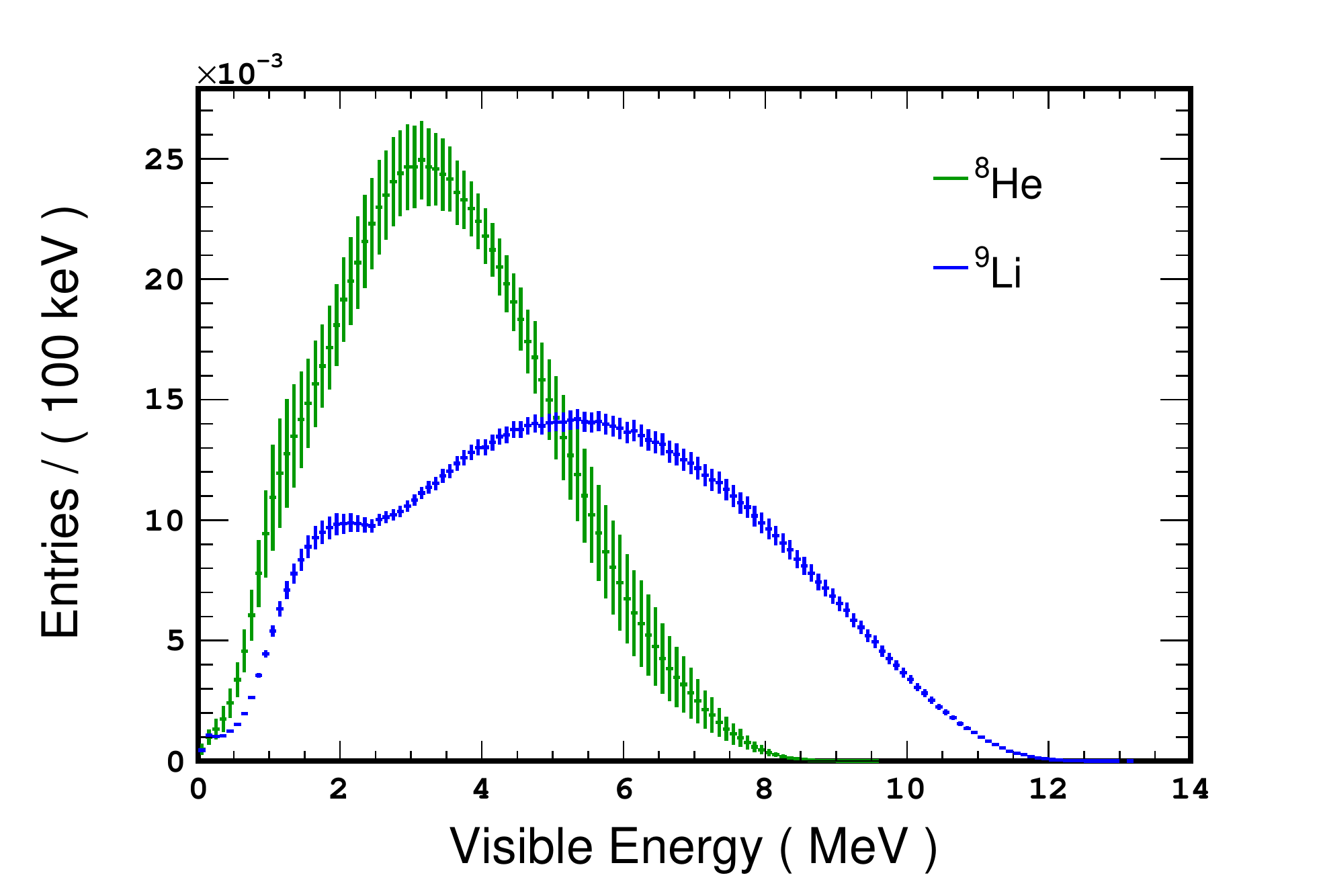}
\caption{\label{fig:mean_MC} Mean simulated $\nucleus{He}{8}$ and $\nucleus{Li}{9}$ spectra. The simulated spectra shown correspond to decays where the emitted neutron's capture has been observed on H or Gd. Each spectrum is normalised to unity. The errors come from the covariance matrices associated with these spectra.}
\end{figure}

The predicted spectra from the error estimator described in section \ref{subsec:error_estimation} are used as inputs in the fit to the data in section \ref{section:Fraction}.
For each radioisotope, and for each neutron capture type, the prediction comes from the sample mean $\overline{\mathbf{b}}$ for the iteration when convergence was reached.
The spectra and covariance matrices are subsequently weighted, taking into account the fractions of Gd ($36\%$) and H ($64\%$) captures observed in the data.
The predicted spectra which are relevant when combining the Gd and H data sets are plotted in figure \ref{fig:mean_MC}.

In the case of $\nucleus{Li}{9}$, the low energy region (at around $\SI{2}{\mega\electronvolt}$) has non--negligible contributions from the strong decays of the levels above the one at $\SI{2.78}{\mega\electronvolt}$ in ${\nucleus{Be}{9}}$. 
The ratios for these decays are little constrained, hence the sizeable uncertainty in that area. 
Above $\SI{4}{\mega\electronvolt}$, electrons start dominating the energy depositions, and the weak magnetism uncertainty takes over as the $\beta$ ratios are well--known. 
Since $\nucleus{He}{8}$ has a lower endpoint than $\nucleus{Li}{9}$, the uncertainty on the $\beta$ branching ratios dominates compared to the weak magnetism uncertainty.
Unfortunately, the $\beta$ ratios feeding the two most populated $\beta\text{-}n$ states in $\nucleus{Li}{8}$ had to be extracted from approximate fits \cite{ref:barker}.
 
\section{Measurements}
\label{section:Measurement}

The selection of a highly pure sample of $\beta\text{-}n$ emitters is explained in the first part of this section.
This sample of events contains a mixture of the cosmogenically produced $\nucleus{Li}{9}$ and $\nucleus{He}{8}$ decays, in unknown quantities.
The second part describes how the predicted spectra are used to separate a background subtracted, and therefore pure sample of $\beta\text{-}n$ emitters into their constituent quantities, denoted by the $\nucleus{He}{8}$ fraction for the ND and FD. 
The third part uses the $\nucleus{He}{8}$ fractions to estimate the total number of $\nucleus{Li}{9}$ and $\nucleus{He}{8}$ produced in each detector from the total measured $\beta\text{-}n$ rates.
The $\nucleus{Li}{9}$ yields are finally compared to other liquid scintillator experiments who have made similar measurements.

\subsection{Selection of \texorpdfstring{$\nucleus{Li}{9}$}{9Li} and \texorpdfstring{$\nucleus{He}{8}$}{8He} candidates}
\label{subsec:sel_9Li_and_8He_cands}

Preliminary selection follows the same criteria as for IBDs published in \cite{ref:DC-GdIII} and \cite{ref:DC-HIII}, corresponding to delayed neutron capture on Gd and H, respectively.
These selections are united to combine the GC and NT volumes used in the analysis.
Prompt candidate selection fell within the visible energy range $0.5\leq E_{p} <\SI{20}{\mega\electronvolt}$ and the delayed neutron within $1.3\leq E_{d}<\SI{10}{\mega\electronvolt}$.
The mean neutron capture time on H and Gd, $\sim\SI{200}{\micro\second}$ and $\sim\SI{30}{\micro\second}$, respectively, determines the selection of candidates which satisfy a time between prompt and delayed event of $ 0.5<\Delta T <\SI{800}{\micro\second}$ and a maximum distance between the two of $\Delta R < \SI{1.2}{\metre}$ is also applied. 
An artificial neural network trained on simulated IBD events, used three variables: $i)$ the time $\Delta T$ and $ii)$ distance $\Delta R$ between prompt and delayed events and $iii)$ the delayed visible energy $E_d$ to further reduce backgrounds \cite{ref:DC-HIII}.
These selection criteria identify coincident signals composed mainly of IBD candidates which are used to measure the neutrino mixing angle $\theta _{13}$. 
However, part of the cosmogenic background can be separated by means of a posterior probability $P$. 
These events removed from the IBD candidates form a relatively pure sample of $\nucleus{Li}{9}$ and $\nucleus{He}{8}$ events which can be used to estimate the $\nucleus{He}{8}$ fraction in section \ref{section:Fraction}.

\begin{figure}[tbp]
	\centering
	\includegraphics[width=0.49\textwidth]{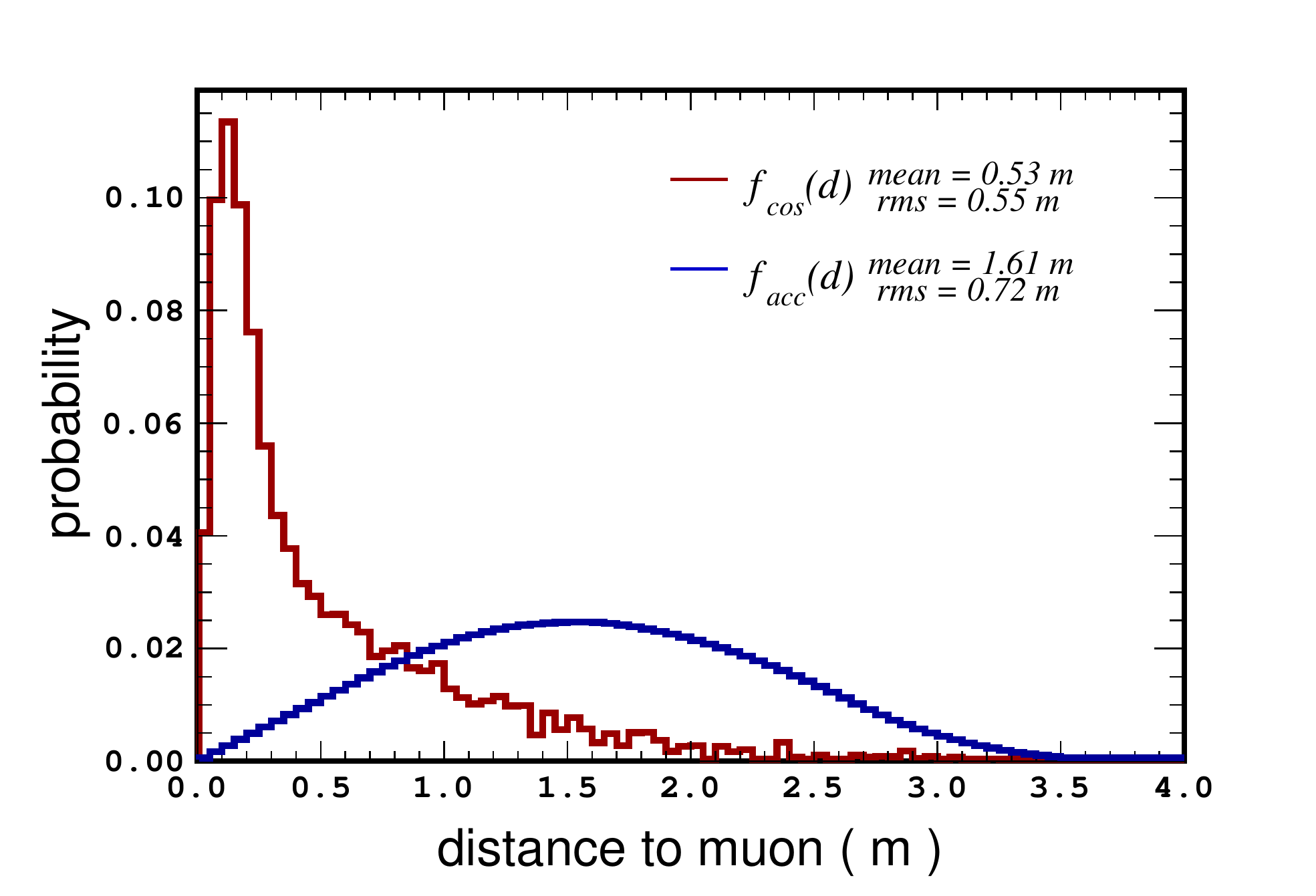}
	\hfill
	\includegraphics[width=0.49\textwidth]{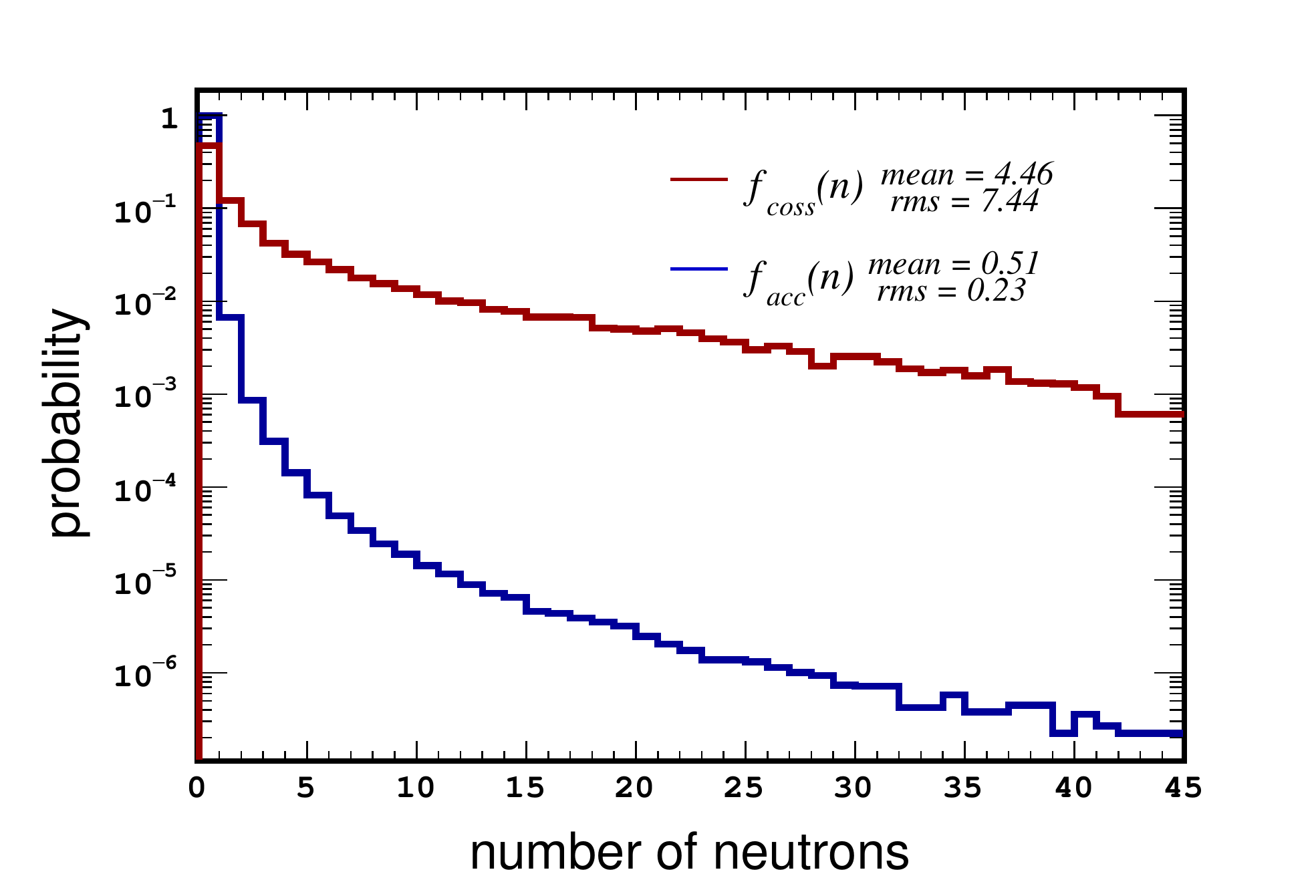}
	\caption{\label{fig:probs} 
	The probability density functions for cosmogenic candidate selection $f_{cos}$ (red) and $f_{acc}$ (blue), used in equation \ref{eq:max_posterior_probability_ratio}.
	Distance between prompt and muon $f(d)$ is shown on the left and number of neutrons after muon $f(n)$ on the right.}
\end{figure}

The probability $P\left(cos \mid n, d\right)$ for a prompt candidate to represent a cosmogenic decay, given that there exists a muon at a distance $d$ from it, which produced $n$ neutrons within $\SI{1}{ms}$, is defined by Bayes' theorem:
\begin{equation}
P \left(cos \mid n, d\right)
= 
\frac{\pi_r \; f_{cos} (n, d) }{\pi_r \; f_{cos} (n, d) + f_{acc} (n, d) }
\,
,
\label{eq:max_posterior_probability_ratio}
\end{equation}
where $f_{cos}$ denotes the joint probability density of $n$ and $d$ for cosmogenics and $f_{acc}$ represents accidental coincidences between IBD candidates and muons.
The probability density functions are shown in figure \ref{fig:probs}.
Including variables other than $d$ and $n$ did not improve the selection efficiency of cosmogenic decays. 
The prior ratio $\pi_r$ is defined as the ratio of the expected cosmogenic rate $r_{cos}$ to the product of the IBD candidate rate $r_{cand}$, the muon rate $r_{\mu}$, and the length $t_{\mathcal{W}}$ of the time window $\mathcal{W}$ used for coincidences:
\begin{equation}
\pi_r = \frac{r_{cos}}{r_{cand} \;  r_{\mu} \; t_{\mathcal{W}}}
\,
.
\label{eq:prior_ratio}
\end{equation}
For $t_{\mathcal{W}} = \SI{700}{ms}$, the ratio $r_{cos}/r_{cand}$ is 0.05 and 0.02, giving prior ratios of $\num{7.7e-3}$ and $\num{5.5e-4}$, for the FD and ND, respectively. 
It is worth stressing that the cosmogenic selection at the ND is all the more difficult because the IBD rate is higher there than at the FD. On the other hand, the increase in the muon rate is largely absorbed by the increase in $r_{cos}$.

The scarcity of $\nucleus{Li}{9}$ and $\nucleus{He}{8}$ events makes them difficult to use to create $f_{cos}$.
Instead, another radioactive isotope $\nucleus{B}{12}$, produced by muons in larger abundance, is utilised.
Although the real distribution of cosmogenic radioisotope production relative to the muon is wider for $\nucleus{B}{12}$, position reconstruction and geometrical effects of the relatively small detector mean that the observed probability densities are similar. 
The so--called on--time window, for the identification of $\nucleus{B}{12}$ correlated events, is shifted by $\SI{1}{ms}$ from muons prior to a prompt candidate; this removes cosmic neutrons, whose capture time is much smaller than the lifetime of the $\beta\text{--}n$ emitters. 
The distribution of the accidental background is determined using several off--time windows which are shifted by more than $\SI{10}{s}$.
Subtracting the average of these from the on--time window they provide $f_{cos}$, normalised to unity they provide $f_{acc}$.
The distributions thus obtained were cross--checked against that of $\nucleus{Li}{9}$+$\nucleus{He}{8}$ and found to agree within the statistical uncertainty.

The posterior probability from \eqref{eq:max_posterior_probability_ratio} is calculated for each prompt candidate $p$ and all the muons $\mu$ in the window $\mathcal{W}_p$ of length $\SI{700}{\milli\second}$ preceding each $p$.
In the $\theta_{13}$ analysis \cite{ref:DC-GdIII,ref:DC-HIII}, the aim was to veto cosmogenic events, dominated by $\beta\text{-}n$ emitters but including to a lesser degree others such as $\nucleus{B}{12}$.
This was achieved by calculating a single (cosmogenic) probability for each prompt by selecting the maximum value as follows:
\begin{equation}
	P_{max} \left(cos\right)
	= \max_{\mu \in \mathcal{W}_p} P \left(cos \mid n_\mu, d_{p\text{-}\mu}\right) \, .
	\label{eq:max_posterior_probability}
\end{equation}
If $P_{max} \left(cos\right) > 0.4$, the prompt event was vetoed as a $\nucleus{Li}{9}$ or $\nucleus{He}{8}$ candidate and used in section \ref{section:Fraction}.

\subsection{\texorpdfstring{$\nucleus{He}{8}$}{8He} fraction measurements}
\label{section:Fraction}

The events selected by $P_{max} \left(cos\right) > 0.4$ in section \ref{subsec:sel_9Li_and_8He_cands} are used to create the data spectra.
The accidental component is estimated using muon--prompt pairs 2--\SI{20}{\second} before the prompt event and then subtracted to give a pure sample of $\nucleus{Li}{9}$ and $\nucleus{He}{8}$ events.
Two--component $\chi ^2$ fits of the data spectra are performed using both time to the previous muon $T$ and prompt visible energy $E_{p}$ information and the number of $\nucleus{Li}{9}$ events $n^{\mathrm{Li}}$ and the number of $\nucleus{He}{8}$ events $n^{\mathrm{He}}$ are left as free parameters.
The results are quoted in terms of the fraction of $\nucleus{He}{8}$ events $f^{He} = n^{\mathrm{He}}/(n^{\mathrm{He}}+n^{\mathrm{Li}})$ and are corrected for time and visible energy cuts. 
The fit is performed separately for the ND and FD as differing overburdens suggest that their $\nucleus{He}{8}$ fractions may not be the same.

A bin size of $1\,{\mathrm{MeV}}$ was used along the energy axis within the range $0.5<E_p<\SI{10.5}{\mega\electronvolt}$ and two bins were used along the time axis, the first $100\,\mathrm{ms}$ and the second $600\,\mathrm{ms}$ long.
The time bins are offset from zero by $1.25\,\mathrm{ms}$ to remove neutrons correlated to muons.
The fitting procedure was tested using toy MC data to choose a bin size which did not bias the results because of the limited statistics and the Gaussian nature of the $\chi^2$ approach.

The $\chi ^2$ is constructed as follows:
\begin{equation}
	\label{eq:ChiDef}
   \chi ^2 = \textbf{y}^{\textbf{T}} \left( \textbf{M}_{stat}^{\,} + \sum_{j}^{\mathrm{H,Gd}} \sum_{c}^{^9\mathrm{Li},^8\mathrm{He}} \textbf{M}_{spec}^{c,j} \right) ^{-1} \textbf{y} \, ,
\end{equation}
where $\textbf{M}_{stat}$ is the covariance matrix corresponding to the statistical uncertainty and
$\textbf{M}_{spec} = \widehat{V} + W$ is the covariance matrix corresponding to the branching ratio and weak magnetism uncertainty from section \ref{subsec:error_estimation}.
$\textbf{y} = y_{(t,e)} - \mu _{(t,e)}$ is the difference between the data and expected value for each bin, where $e$ corresponds to the bin number along the energy axis and $t$ along the time axis.
The expected value for each bin is evaluated as follows:
\begin{equation}
	\label{eq:meanval}
	\mu _{(t,e)} = \sum ^{\mathrm{H,Gd}}_{j} \sum _{c}^{^9\mathrm{Li},^8\mathrm{He}}  F_j n^{c} \left( e^{-\frac{  T_t }{\tau _{c}}} - e^{-\frac{  T_{t+1} }{\tau _{c}}} \right ) S(E_{p})^{c,j}_e \, ,
\end{equation}
where $F_j$ is the fraction of events where the neutron was captured on H or Gd, $c$ is the cosmogenic radioisotope $\nucleus{Li}{9}$ or $\nucleus{He}{8}$, 
$\tau _c$ is the lifetime $\SI{257}{{\milli\second}}$ for $\nucleus{Li}{9}$ and $\SI{172}{\milli\second}$ for $\nucleus{He}{8}$, 
$T_t$ is the start of the time bin $t$, and $S(E_{\text{p}})^{c,j}_e$ is the fraction of the predicted spectrum expected in bin $e$.
The effect of the energy scale uncertainty on the fit results was tested by altering the spectra by $\pm1\sigma$ uncertainty.
It was found to be negligible as a result of the much larger statistical uncertainty. 

\begin{figure}[tbp]
	\centering
	\includegraphics[width=0.49\textwidth]{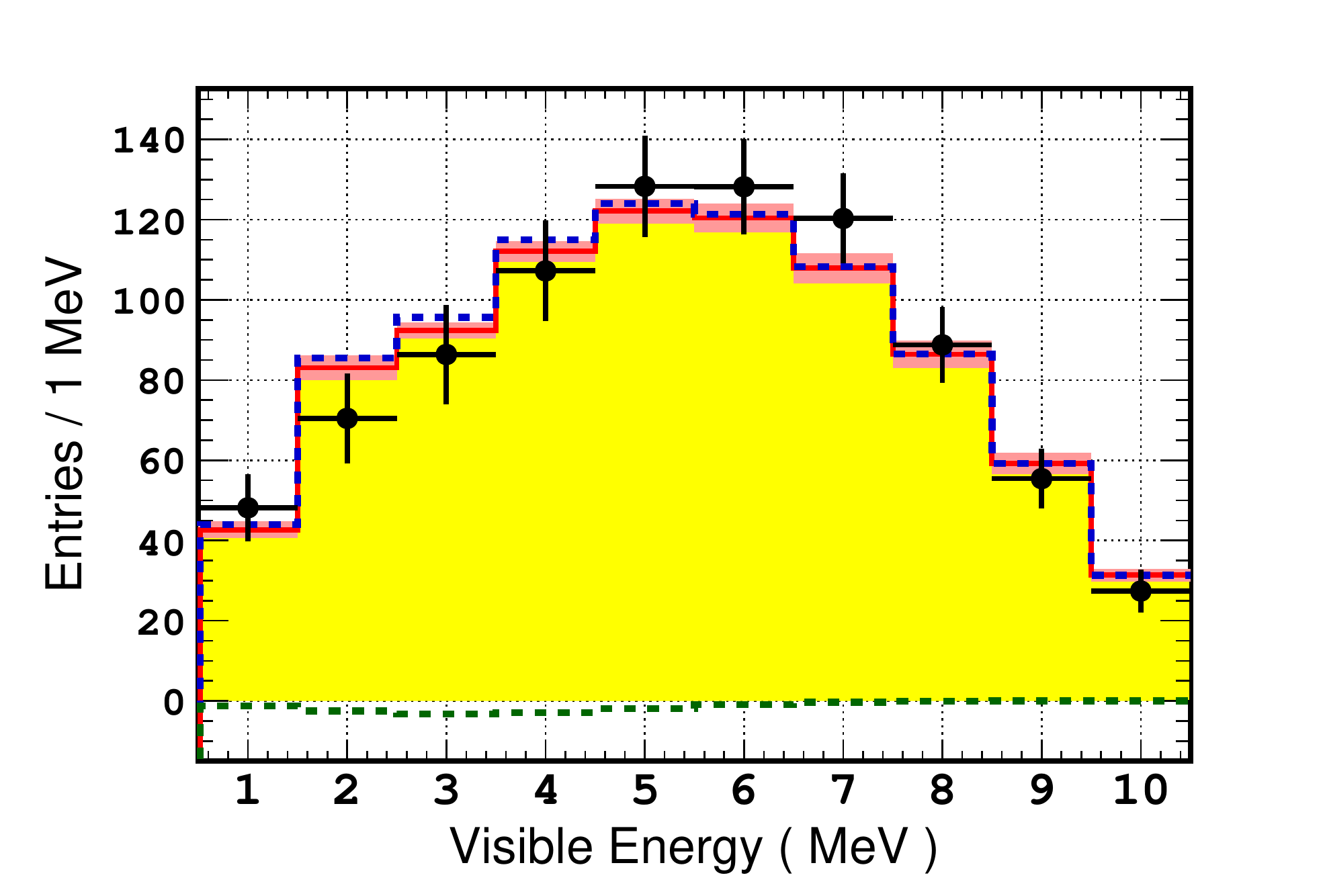}
	\hfill
	\includegraphics[width=0.49\textwidth]{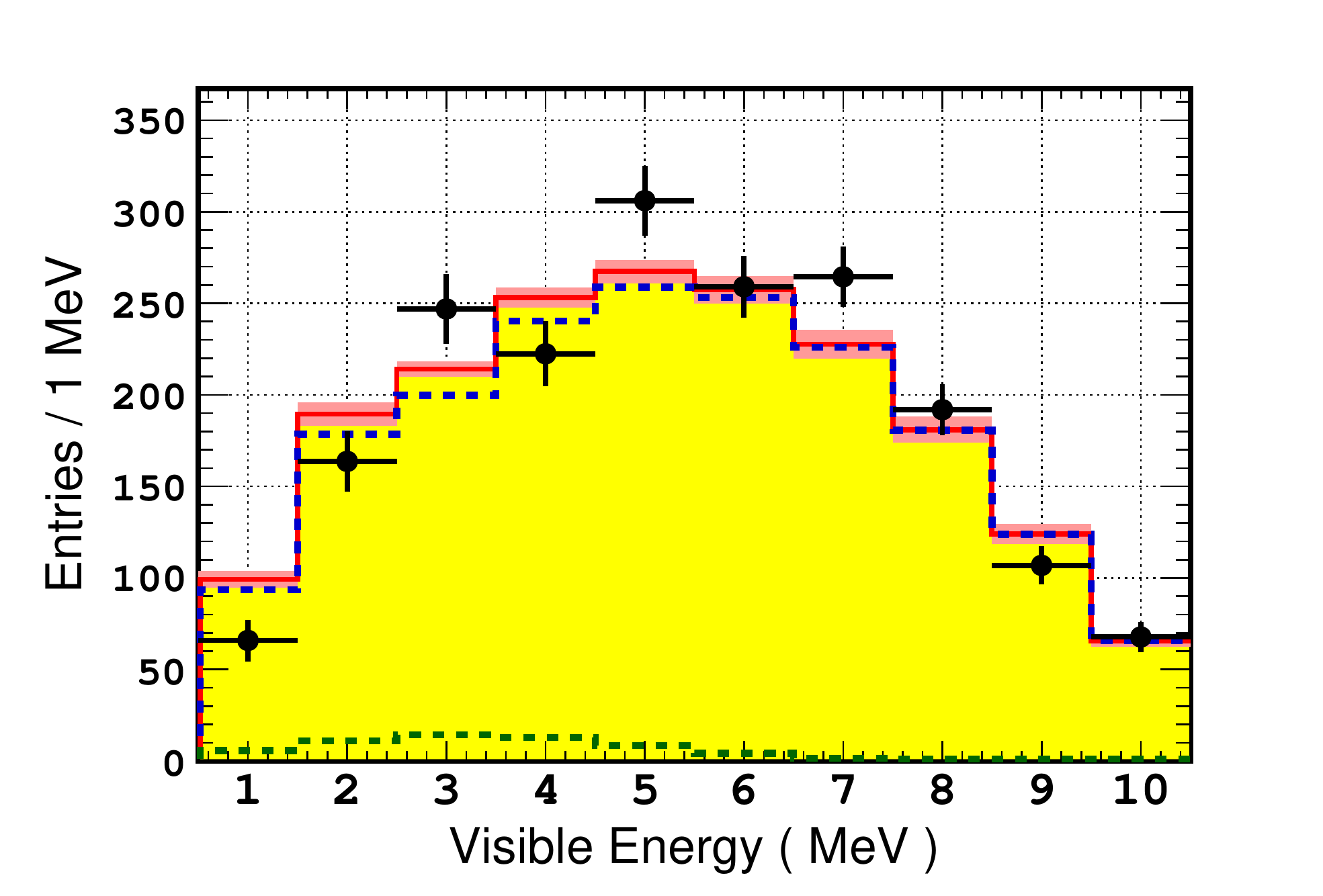}
	\caption{\label{fig:BestFitSpecs} 
	The ND (left) and FD (right) visible energy spectra of the $\nucleus{Li}{9}$+$\nucleus{He}{8}$ data (black points). 
	Overlain are the best fit results with $\nucleus{Li}{9}$ (blue dashed line), $\nucleus{He}{8}$ (green dashed line), and their sum (solid red line).
	The pink error bands represent the uncertainty on the predicted spectrum.
	The goodness of the fits are $\chi ^2 /d.o.f = 16.5/18$ (ND) and $\chi ^2 /d.o.f = 33.5/18$ (FD).}
\end{figure}

For the ND, $f^{He}_{ND} = (-1.5\pm4.5)\%$ with $\chi ^2 /d.o.f = 16.5/18$ giving a probability $p=0.56$.
For the FD, $f^{He}_{FD} = (2.9\pm3.5)\%$ with $\chi ^2 /d.o.f = 33.5/18$ giving $p=0.01$.
Both results return a $\nucleus{He}{8}$ component compatible with zero.
The fit results can be seen as a function of $E_{\text{vis}}$, summed along the time axis in figure \ref{fig:BestFitSpecs} for the ND (left) and FD (right).
The pink error bands represent the uncertainty on the predicted spectrum.

\subsection{Cosmogenic yields, production rates, and cross sections}
\label{section:yields}

Constraints on the $\nucleus{He}{8}$ fractions from the previous section allow the yields, production rates, and cross sections of $\nucleus{Li}{9}$ and $\nucleus{He}{8}$ to be quoted separately.
As the ND, FD, Borexino, and KamLAND detectors all have distinct overburdens they are subject to different muon spectra represented by the mean muon energy $<E_{\mu}>$.
By combining measurements of the $\nucleus{Li}{9}$ and $\nucleus{He}{8}$ yields at all these sites with $<E_{\mu}>$ a relationship between the two can be constrained.
This can in turn be used to predict the background rates of $\nucleus{Li}{9}$ and $\nucleus{He}{8}$ at future liquid scintillator experiments.

To estimate the yields, production rates and cross sections for DC, the $\beta\text{-}n$ rates or $r_{\beta\text{-}n}$ are determined first. 
These are a combination of $\nucleus{Li}{9}$ and $\nucleus{He}{8}$ events as selection cannot differentiate between the two.
Selection is the same as described in section \ref{subsec:sel_9Li_and_8He_cands}, except that the minimum prompt energy is $E_{p} > \SI{1}{\mega\electronvolt}$ for reasons discussed in \cite{ref:DC-HIII} and $r_{\beta\text{-}n}$ is determined separately for $P_{max}<0.4$ and $P_{max}>0.4$ as calculated using equation \ref{eq:max_posterior_probability}.
The latter category is dominated by $\beta\text{-}n$ events and is determined by subtracting the average number of events in twelve off-time windows from an on-time window, where each window is $\SI{700}{\milli\second}$ long. 
The measured rates can be found in table \ref{table:liherates}. The former category is dominated by IBDs, but includes some accidentals and fast neutrons. 
A complex analysis to determine $\theta_{13}$ is implemented which incorporates the aforementioned backgrounds to a fit of $E_{p}$, from which the $\beta\text{-}n$ rate is determined as a by-product.
These rates can be found in table \ref{table:liherates} and were cross-checked by applying a fit to the distribution of time differences between prompt events and previous muons giving $12.32\pm \SI{2.01}{day^{-1}}$ and $3.01\pm \SI{0.60}{day^{-1}}$ for the ND and FD, respectively.
Also given in table \ref{table:liherates} are the resulting total $\beta\text{-}n$ rates along with the selection efficiencies which apply to both $\nucleus{He}{8}$ and $\nucleus{Li}{9}$.

The $\nucleus{He}{8}$ fraction measurements from section \ref{section:Fraction} are used to estimate the corresponding rates $r_c$ of each cosmogenic radioisotope $c$ by splitting the efficiency corrected total $\beta\text{-}n$ rates into their respective $\nucleus{He}{8}$ and $\nucleus{Li}{9}$ components, after which they are further corrected for cosmogenic dependent efficiencies.
The $\nucleus{He}{8}$ fraction measurements are assumed to be the same for both $P_{max}<0.4$ and $P_{max}>0.4$.
\begin{table}[tbp] \centering

\begin{tabular}{| l l |  c  c |} \hline
																					&									&	ND 						&  FD 				\\ \hline
\multicolumn{2}{|c|}{livetime (days)}																		& 	257.96 					&	818.18			\\
\multicolumn{2}{|c|}{$\epsilon_{\beta\text{-}n} $	(\%)}												&	$81.7\pm0.1$			& $84.2\pm0.2$		\\ \hline
\multirow{3}{*}{$r_{\beta\text{-}n}$ ($\SI{}{day^{-1}}$)}		& $P_{max}<0.4$  				& $14.52\pm1.48$			& $2.62\pm0.27$	\\
																					& $P_{max}>0.4$				&	$3.99\pm0.14$			& $2.81\pm0.07$	\\
																					& Total 							&	$18.51\pm1.49$			& $5.43\pm0.28$	\\
\hline
\multicolumn{2}{|c|}{$r_{Li}$ ($\SI{}{day^{-1}}$)}	  													& $22.65\pm2.08$			& $6.25\pm0.40$	\\ 
\multicolumn{2}{|c|}{$r_{He}$ ($\SI{}{day^{-1}}$)}	  													& $<6.24$					& $0.19\pm0.40$	\\ 
\hline
\end{tabular}
\caption{\label{table:liherates} The livetime, $\beta\text{-}n$ rates along with their efficiencies $\epsilon_{\beta\text{-}n}$, and the efficiency corrected total $\beta\text{-}n$ rate separated into the individual cosmogenic rates $r_{c}$ used to estimate the yields, production rates, and cross sections. 
}
\end{table}

The yield is defined in \cite{ref:KamLANDCosmo,ref:BorexinoCosmo} as follows:
\begin{equation}
	\label{eq:yield1}
	Y_c = \frac{\mathcal{N}_c}{ R_{\mu} T_{L} \cdot \langle L_{\mu} \rangle \cdot \rho } \, , 
\end{equation}
and the production rate as:
\begin{equation}
	\label{eq:prodrate}
	R_c = \frac{\mathcal{N}_c}{ V \cdot \rho \cdot T_{L} } \, , 
\end{equation}
where $\mathcal{N}_c = (r_c \cdot T_{L})/\epsilon _c$ is the total number of the cosmogenic radioisotope $c$ created in the total liquid scintillator volume $V = 32.8\,\mathrm{m^3}$, $\epsilon _c$ is the product of all efficiencies specific to that radioisotope which is dominated by the non $\beta\text{-}n$ branching ratios (\SI{49}{\percent} for $\nucleus{Li}{9}$ and \SI{84}{\percent} for $\nucleus{He}{8}$), $R_{\mu}$ is the muon rate in $V$, $\langle L_{\mu} \rangle$ is the average muon track length in the volume, $T_L$ is the live time, the density $\rho=\SI{0.8}{g.cm^{-3}}$, and $r_c$, shown in table \ref{table:liherates}, is the rate of either $\nucleus{Li}{9}$ or $\nucleus{He}{8}$ separated from the total $\beta\text{-}n$ rate using the $^8$He fraction measured in section \ref{section:Fraction}.
As the muon flux can be defined as $\phi_{\mu} = R_{\mu} \langle L_{\mu} \rangle / V$ a substitution can be made into equation \ref{eq:yield1} to give a new definition of the yield as:
\begin{equation}
	\label{eq:yield2}
	Y_c = \frac{\mathcal{N}_c}{ \phi_{\mu} \cdot V \cdot T_{L} \cdot \rho }, 
\end{equation}
where the muon fluxes are $3.64\pm\SI{0.04}{m^{-2}.s^{-1}}$ and $0.70\pm\SI{0.01}{m^{-2}.s^{-1}}$ for the ND and FD, respectively, measured using data in \cite{ref:DC-MuMod}. 
The production cross section can be inferred from the yield via the relation:
\begin{equation}
	\label{eq:yieldCrossSec}
	Y_c = \frac{\sigma_{c}}{m_{T}} \, ,
\end{equation}
where $m_T$ is the mass of the target atom, in this case $\nucleus{C}{12}$. 
It should be noted that this cross section is averaged over the muon energy spectrum corresponding to the specific detector depth.
The yields, production rates, and cross sections for the ND and FD are given in table \ref{table:yield}.
\begin{table}[tbp] \centering
\begin{tabular}{| l c | c c c c c |}
 \hline 
							  &								& $< E_{\mu} >$ 						& $c$ 		& $Y$																 	& $R$ 									&	$\sigma$							\\
							  &								& $ (\mathrm{GeV})$					&				& $(\times10^{-8}\,\mu^{-1}\mathrm{g^{-1}cm^{2}})$		& $(\mathrm{kton^{-1}d^{-1}})$   & 	$\mathrm{(\mu barns)}$		\\
 \hline
 \multirow{4}{*}{DC}   & \multirow{2}{*}{ ND}	& \multirow{2}{*}{$32.1\pm2.0$}	& $^9$Li 	& $5.51\pm0.52$ 													& $1733\pm161$							&	$1.10\pm0.10$	  		\\
							  & 								&											& $^8$He 	& $<4.96$ 															& $<1561$ 								&	$<0.99$   				\\ 
							  & \multirow{2}{*}{ FD}	&	\multirow{2}{*}{$63.7\pm5.5$} & $^9$Li 	& $7.90\pm0.51$ 													& $478\pm31$ 							&	$1.57\pm0.10$			\\
							  & 								&											& $^8$He 	& $0.77\pm1.61$ 													& $47\pm98$ 							&	$0.15\pm0.32$			\\ 
 \hline 
 \multirow{3}{*}{Daya Bay}   & EH1					&	$57$									& $^9$Li 	& $7.66\pm0.80$ 													& $-$ 									&	$-$						\\
									  & EH2					&	$58$									& $^9$Li 	& $7.72\pm0.91$ 													& $-$ 									&	$-$						\\
	 								  & EH3					&	$137$									& $^9$Li 	& $15.65\pm1.85$ 													& $-$ 									&	$-$						\\
 \hline 
 \multicolumn{2}{|l|}{\multirow{2}{*}{KamLAND}}	& \multirow{2}{*}{$260\pm8$}		& $^9$Li 	& $22\pm2$ 															& $2.8\pm0.2$							&	$-$	  					\\
							&									&											& $^8$He 	& $7\pm4$ 															& $1.0\pm0.5$ 							&	$-$   					\\ 
 \hline 
 \multicolumn{2}{|l|}{\multirow{2}{*}{Borexino}}& \multirow{2}{*}{$283\pm19$}	   & $^9$Li 	& $29\pm3$ 															& $0.83\pm0.09$						&	$-$	  					\\
						&										&											& $^8$He 	& $<15$ 																& $<0.42$ 								&	$-$   					\\ 
 \hline 

\end{tabular} 
\caption{\label{table:yield} The yields $Y$, production rates $R$, and cross sections for $\nucleus{Li}{9}$ and $\nucleus{He}{8}$ produced in the ND and FD along with the corresponding mean muon energy $<E_{\mu}>$. 
In the ND case the $\nucleus{Li}{9}$ values are calculated for $f^{He}=0\%$ and a $3\sigma$ upper limit is given for the $\nucleus{He}{8}$ values.
The same values are given for Daya Bay \cite{ref:DB2017,ref:DBMuSys}, KamLAND \cite{ref:KamLANDCosmo}, and Borexino \cite{ref:BorexinoCosmo} where available and any limits are given at $3\sigma$.
The Daya Bay $^9$Li yields are calculated under the assumption that the measurements of their $\beta\text{-}n$ emitters are composed purely of $^9$Li.}
\end{table}

\begin{figure}[tbp]
	\centering
	\includegraphics[width=0.7\textwidth]{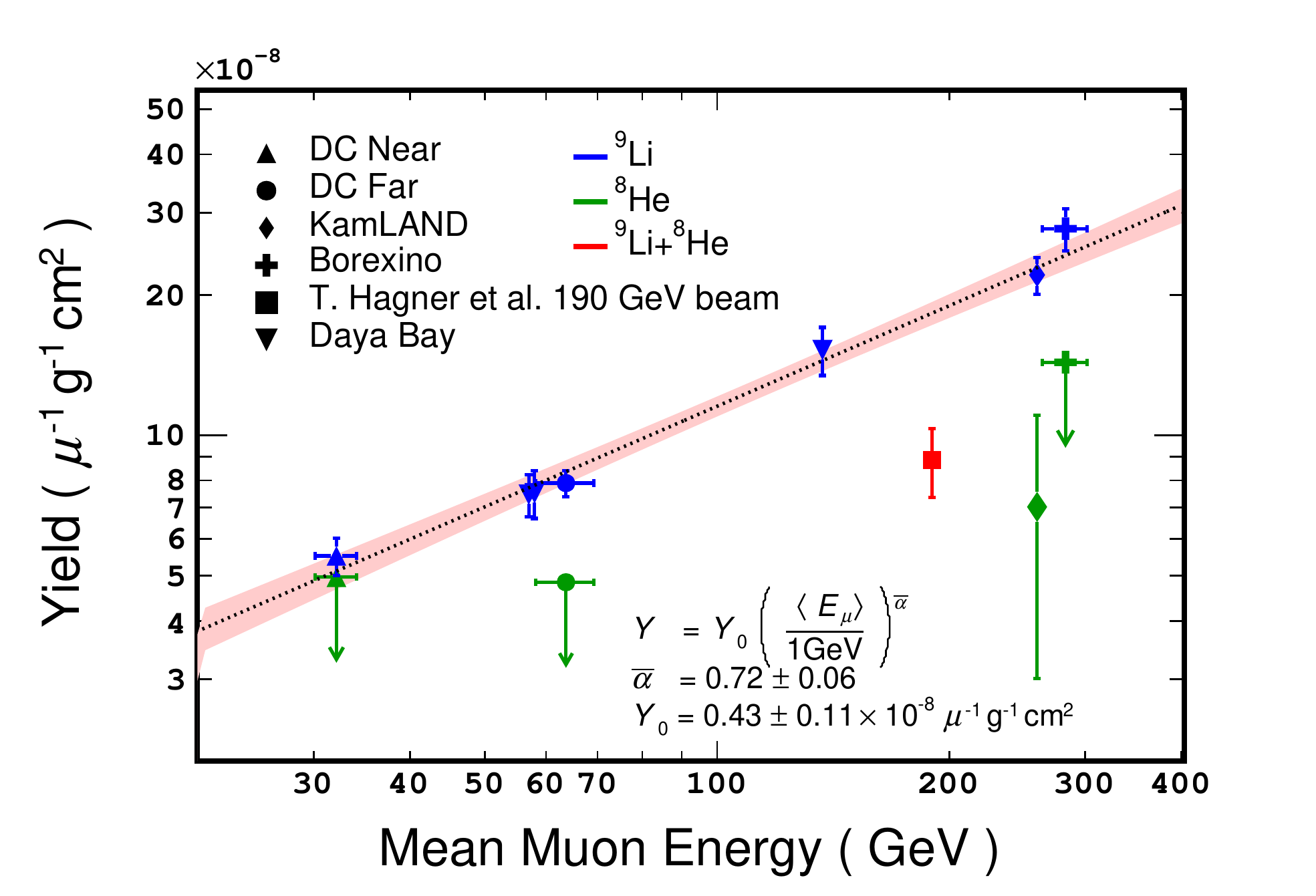}
	\caption{\label{fig:yields}The ND and FD yields shown separately for $\nucleus{Li}{9}$ and $\nucleus{He}{8}$ as a function of the mean muon energy $\langle E_{\mu} \rangle$, alongside those from KamLAND, Borexino, and Daya Bay. 
	The arrows depict a 3$\sigma$ upper limit and their lengths equal $1\sigma$. 
	A fit to the $\nucleus{Li}{9}$ yields of the ND, FD, Borexino, and KamLAND returns the power law exponent $\overline{\alpha} = 0.72\pm0.06$ with $\chi ^2 / d.o.f = 4.0/2$.} 
\end{figure}

As discussed in \cite{ref:Hagner2000}, the yields follow a dependence on the mean muon energy $E_{\mu}$ according to a power exponent $\alpha$ as $Y \propto E_{\mu} ^{\alpha}$.
In the case of underground sites the yield is given as a function of the mean muon energy which can be fitted with:
\begin{equation}
	Y = Y_0 \left ( \frac{\langle E_{\mu} \rangle}{\SI{1}{GeV}} \right )^{\overline{\alpha}} \, , \label{eq:powlaw}
\end{equation}
where $\overline{\alpha}$ is used to denote the power law as a function of $\langle E_{\mu} \rangle$ instead of a mono--energetic muon energy.
The mean muon energies at the ND and FD are $32.1\pm \SI{2.0}{\giga\electronvolt}$ and $63.7\pm \SI{5.5}{\giga\electronvolt}$, respectively, evaluated with a dedicated MUSIC simulation \cite{ref:music}.
The yields can be compared to measurements by KamLAND \cite{ref:KamLANDCosmo}, Borexino \cite{ref:BorexinoCosmo}, and Daya Bay \cite{ref:DB2017,ref:DBMuSys} as demonstrated in figure \ref{fig:yields}, where the values have been corrected to represent the carbon density of the DC detectors.
The Daya Bay yields are estimated from their $\beta \text{-}n$ rates at each experimental hall assuming they contain $100\%$ $\nucleus{Li}{9}$.
As such, and because not all of the uncertainties and efficiencies are known, they are omitted from the fit for $\overline{\alpha}$.

Also included in the figure is a measurement at CERN using the SPS muon beam aimed at a liquid scintillator target, where the combined cross section for $\nucleus{Li}{9}$+$\nucleus{He}{8}$ is shown converted into the yield using equation \ref{eq:yieldCrossSec} and corrected by a factor of $0.87\pm0.03$ to take into account the energy spectrum of cosmic muons \cite{ref:Hagner2000}.
The measurement displayed by the red marker clearly underestimates the expected yield as defined by the best fit line in figure \ref{fig:yields}.
The production rates for $\nucleus{C}{11}$, $\nucleus{Li}{8}$, $\nucleus{B}{8}$ and $\nucleus{He}{6}$ predicted by \cite{ref:Hagner2000} for the Borexino experiment were also lower than subsequent measurements made by Borexino \cite{ref:BorexinoCosmo}.
This could be due to an underestimation of the efficiency correction for radioisotope production at larger distances from the muon beam.
\label{section:Yields}

A covariance matrix, separating the uncertainties on the yields and mean muon energies between the ND and FD into their correlated and uncorrelated parts, is used when performing the fit to the $\nucleus{Li}{9}$ yields, giving $\overline{\alpha} = 0.72\pm0.06$ and $Y_0 = (0.43\pm0.11)\times 10^{-8} \mathrm{\mu ^{-1} g^{-1} cm ^{2}}$ with a correlation of $\rho = -0.98$.
The minimum was found at $\chi ^2 / d.o.f = 4.0/2$, demonstrating the reliability of the model over these detector depths.
No such fit was performed for $\nucleus{He}{8}$ as the values are compatible with zero.

The power law relationship described by equation \ref{eq:powlaw} can be used to estimate the $\nucleus{Li}{9}$ yield for future liquid scintillator experiments.
With a mean muon energy of \SI{215}{\giga\electronvolt}, the $\nucleus{Li}{9}$ yield for JUNO is estimated to be $(19.96\pm1.21)\times 10^{-8} \mathrm{\mu ^{-1} g^{-1} cm ^{2}}$. 
This does not include the uncertainty on JUNO's mean muon energy.

\section{Conclusion}
\label{section:Conclusion}
The long--lived $\beta\text{-}n$ emitters produced in the Double Chooz ND and FD by spallation interactions of high--energy muons satisfy the time and energy selection criteria of IBD candidates.
The distance between the muon track and prompt signal of the IBD along with the neutron multiplicity following that muon can be used to effectively select a pure sample of cosmogenic events.

The two components expected to form the selected $\beta\text{-}n$ emitters of the ND and FD are $\nucleus{Li}{9}$ and to a smaller extent $\nucleus{He}{8}$. 
Thanks to accurate simulations of the complex decays schemes of these radioisotopes and the following detector response to the complete set of finals states, a fit of these two components could be performed.  
A strong constraint has been set on the production rate of the $\nucleus{He}{8}$ radioisotope, found to be compatible with zero at both detector sites.

A clear relationship between the production yield of the $\nucleus{Li}{9}$ radioisotope with $\left<E_\mu\right>$ has been measured between the two detector sites attributed to the different muon spectra reaching the fiducial volume ($32.1\pm\SI{2.0}{\giga\electronvolt}$ for the ND and $63.7\pm\SI{5.5}{\giga\electronvolt}$ for the FD). 
These results extend the study of $\nucleus{Li}{9}$ production towards smaller mean muon energies than other published experiments. 
Combining all available data a simple power law (equation \ref{eq:powlaw}) is found to describe the $\nucleus{Li}{9}$ production yield with good accuracy over one order of magnitude in the muon energy and does not match the independent measurement performed with a 190 MeV monochromatic muon beam.
This power law provides a new reference for future liquid scintillator based reactor neutrino experiments.


\acknowledgments
We thank the French electricity company EDF;
the European fund FEDER;
the R\'egion de Champagne Ardenne;
the D\'epartement des Ardennes;
and the Communaut\'e de Communes Ardenne Rives de Meuse.
We acknowledge the support of
the CEA, CNRS/IN2P3, the computer centre CC-IN2P3,
and
LabEx UnivEarthS in France;
the Max Planck Gesellschaft,
the Deutsche Forschungsgemeinschaft DFG,
the Transregional Collaborative Research Center TR27,
the excellence cluster ``Origin and Structure of the Universe'',
and
the Maier-Leibnitz-Laboratorium Garching in Germany;
the Ministry of Education, Culture, Sports, Science and Technology of Japan (MEXT),
and
the Japan Society for the Promotion of Science (JSPS) in Japan;
the Ministerio de Econom\'ia, Industria y Competitividad (SEIDI-MINECO) under grants FPA2016-77347-C2-1-P and MdM-2015-0509 in Spain;
the Department of Energy and the National Science Foundation,
and
Department of Energy in the United States;
the Russian Academy of Science,
the Kurchatov Institute,
and
the Russian Foundation for Basic Research (RFBR) in Russia;
the Brazilian Ministry of Science, Technology and Innovation (MCTI),
the Financiadora de Estudos e Projetos (FINEP),
the Conselho Nacional de Desenvolvimento Cient\'ifico e Tecnol\'ogico (CNPq),
the S\~ao Paulo Research Foundation (FAPESP),
and
the Brazilian Network for High Energy Physics (RENAFAE) in Brazil.



\appendix
\section{Explicit decay chains}
\label{sec:explicit_decay_chains}
The explicit decay chains after the $\beta$--decays of $\nucleus{Li}{9}$ and $\nucleus{He}{8}$ may be found in table \ref{table:9Li_decays} and table \ref{table:8He_decays}, respectively.

\begin{table}[!htbp] \centering
\begin{tabular}{lcccc}
	$\nucleus{Be}{9}^{2.43} $	& $\longrightarrow$ & $\nucleus{Be}{8}$ 						& $\longrightarrow$ & $\left(\alpha, \alpha\right)$\\
	$\nucleus{Be}{9}^{2.43} $	& $\longrightarrow$ & $\nucleus{He}{5}$ 						& $\longrightarrow$ & $\left(\alpha, n\right)$		\\
	$\nucleus{Be}{9}^{2.43} $	& $\longrightarrow$ & $\left(\alpha, \alpha, n\right)$	\\
	$\nucleus{Be}{9}^{2.78} $	& $\longrightarrow$ & $\nucleus{Be}{8}$ 						& $\longrightarrow$ & $\left(\alpha, \alpha\right)$\\
	$\nucleus{Be}{9}^{2.78} $	& $\longrightarrow$ & $\nucleus{Be}{8}^{3.03}$ 				& $\longrightarrow$ & $\left(\alpha, \alpha\right)$\\
	$\nucleus{Be}{9}^{2.78} $	& $\longrightarrow$ & $\nucleus{He}{5}$ 						& $\longrightarrow$ & $\left(\alpha, n\right)$		\\
	$\nucleus{Be}{9}^{2.78} $	& $\longrightarrow$ & $\left(\alpha, \alpha, n\right)$	\\
	$\nucleus{Be}{9}^{7.94} $	& $\longrightarrow$ & $\nucleus{Be}{8}$						& $\longrightarrow$ & $\left(\alpha, \alpha\right)$\\
	$\nucleus{Be}{9}^{7.94} $	& $\longrightarrow$ & $\nucleus{Be}{8}^{3.03}$ 				& $\longrightarrow$ & $\left(\alpha, \alpha\right)$\\
	$\nucleus{Be}{9}^{7.94} $	& $\longrightarrow$ & $\nucleus{He}{5}$ 						& $\longrightarrow$ & $\left(\alpha, n\right)$		\\
	$\nucleus{Be}{9}^{7.94} $	& $\longrightarrow$ & $\nucleus{He}{5}^{3.74}$ 				& $\longrightarrow$ & $\left(\alpha, n\right)$		\\
	$\nucleus{Be}{9}^{7.94} $	& $\longrightarrow$ & $\left(\alpha, \alpha, n\right)$	\\
	$\nucleus{Be}{9}^{11.28}$ 	& $\longrightarrow$ & $\nucleus{Be}{8}$ 						& $\longrightarrow$ & $\left(\alpha, \alpha\right)$\\
	$\nucleus{Be}{9}^{11.28}$ 	& $\longrightarrow$ & $\nucleus{Be}{8}^{11.35}$ 			& $\longrightarrow$ & $\left(\alpha, \alpha\right)$\\
	$\nucleus{Be}{9}^{11.28}$ 	& $\longrightarrow$ & $\nucleus{Be}{8}^{3.03}$ 				& $\longrightarrow$ & $\left(\alpha, \alpha\right)$\\
	$\nucleus{Be}{9}^{11.28}$ 	& $\longrightarrow$ & $\nucleus{He}{5}$ 						& $\longrightarrow$ & $\left(\alpha, n\right)$		\\
	$\nucleus{Be}{9}^{11.28}$ 	& $\longrightarrow$ & $\nucleus{He}{5}^{3.74}$ 				& $\longrightarrow$ & $\left(\alpha, n\right)$		\\
	$\nucleus{Be}{9}^{11.28}$ 	& $\longrightarrow$ & $\left(\alpha, \alpha, n\right)$	\\
	$\nucleus{Be}{9}^{11.81}$ 	& $\longrightarrow$ & $\nucleus{Be}{8}$ 						& $\longrightarrow$ & $\left(\alpha, \alpha\right)$\\
	$\nucleus{Be}{9}^{11.81}$ 	& $\longrightarrow$ & $\nucleus{Be}{8}^{11.35}$ 			& $\longrightarrow$ & $\left(\alpha, \alpha\right)$\\
	$\nucleus{Be}{9}^{11.81}$ 	& $\longrightarrow$ & $\nucleus{Be}{8}^{3.03}$ 				& $\longrightarrow$ & $\left(\alpha, \alpha\right)$\\
	$\nucleus{Be}{9}^{11.81}$ 	& $\longrightarrow$ & $\nucleus{He}{5}$ 						& $\longrightarrow$ & $\left(\alpha, n\right)$		\\
	$\nucleus{Be}{9}^{11.81}$ 	& $\longrightarrow$ & $\nucleus{He}{5}^{3.74}$ 				& $\longrightarrow$ & $\left(\alpha, n\right)$		\\
	$\nucleus{Be}{9}^{11.81}$ 	& $\longrightarrow$ & $\left(\alpha, \alpha, n\right)$	\\
	\end{tabular}
	\caption{\label{table:9Li_decays} $\nucleus{Be}{9}$ decay paths relevant after the $\beta$--decay of $\nucleus{Li}{9}$. 
	The energy levels (in \si{MeV}) relative to the ground state of $\nucleus{Be}{9}$ are indicated as superscripts of the chemical elements; when no superscript is present, the ground state of the considered nucleus must be understood.}
\end{table}

\begin{table}[!htbp] \centering
	\begin{tabular}{lcccc}
	$\nucleus{Li}{8}^{3.21}$ 	& $\longrightarrow$ & $\nucleus{Li}{7}$				\\
	$\nucleus{Li}{8}^{3.21}$ 	& $\longrightarrow$ & $\nucleus{Li}{7}^{2.51}$ 		& 	$\longrightarrow$ & $\left(\nucleus{Li}{7}, \gamma\right)$\\
	$\nucleus{Li}{8}^{3.21}$ 	& $\longrightarrow$ & $\nucleus{Li}{7}^{2.51}$		\\
	$\nucleus{Li}{8}^{5.4} $	& $\longrightarrow$ & $\nucleus{Li}{7}$				\\
	$\nucleus{Li}{8}^{5.4} $	& $\longrightarrow$ & $\nucleus{Li}{7}^{2.51}$	 	& 	$\longrightarrow$ & $\left(\nucleus{Li}{7}, \gamma\right)$\\
	$\nucleus{Li}{8}^{5.4} $	& $\longrightarrow$ & $\nucleus{Li}{7}^{2.51}$		\\
	$\nucleus{Li}{8}^{5.4} $	& $\longrightarrow$ & $\left(\alpha, t, n\right)$	\\
	$\nucleus{Li}{8}^{9.67}$ 	& $\longrightarrow$ & $\nucleus{He}{5}$ 				& $\longrightarrow$ & $\left(\alpha, n\right)$	\\
	$\nucleus{Li}{8}^{9.67}$ 	& $\longrightarrow$ & $\nucleus{Li}{7}$				\\
	$\nucleus{Li}{8}^{9.67}$ 	& $\longrightarrow$ & $\nucleus{Li}{7}^{2.51}$ 		& 	$\longrightarrow$ & $\left(\nucleus{Li}{7}, \gamma\right)$	\\
	$\nucleus{Li}{8}^{9.67}$ 	& $\longrightarrow$ & $\nucleus{Li}{7}^{2.51}$		\\
	$\nucleus{Li}{8}^{9.67}$ 	& $\longrightarrow$ & $\left(\alpha, t, n\right)$	\\
	\end{tabular}
	\caption{\label{table:8He_decays} $\nucleus{Li}{8}$ decay paths relevant after the $\beta$--decay of $\nucleus{He}{8}$. 
	The energy levels (in \si{MeV}) relative to the ground state of $\nucleus{Li}{8}$ are indicated as superscripts of the chemical elements; when no superscript is present, the ground state of the considered nucleus must be understood.}
\end{table}

\end{document}